\documentclass[conference]{IEEEtran}

\usepackage{cite}
\usepackage[pdftex]{graphicx}

\usepackage{tikz}
\usepackage{amsmath}
\usepackage{multicol}
\usepackage{multirow}
\usepackage{fontawesome}
\usepackage{array}
\usepackage{mathtools}
\usepackage{colortbl}
\usepackage{xspace}
\usepackage{kmath}
\usepackage{booktabs}
\usepackage[hidelinks]{hyperref}
\usepackage{pgfplots}
\pgfplotsset{compat=1.18}
\definecolor{mygray}{HTML}{ededed}
\usepackage{longtable}
\usepackage{subcaption}
\usepackage{graphicx}

\begin{document}

\title{Exploring Older Adults' Perceptions and Experiences with Online Dating}

% \author{\IEEEauthorblockN{Anonymous}}

% \author{\IEEEauthorblockN{Musan Fatima}
% \IEEEauthorblockA{\textit{University of Denver}\\
% muskan.fatima@du.edu}
% \and
% \IEEEauthorblockN{Naheem Noah}
% \IEEEauthorblockA{\textit{University of Denver}\\
% naheem.noah@du.edu}
% \and
% \IEEEauthorblockN{Sanchari Das}
% \IEEEauthorblockA{\textit{University of Denver}\\
% sanchari.das@du.edu}
% }
% \author{\IEEEauthorblockN{Muskan Fatima, Naheem Noah, Sanchari Das} \IEEEauthorblockA{\textit{University of Denver}\\{\{firstname.lastname\}}@du.edu} }

\author{
    \IEEEauthorblockN{Muskan Fatima\textsuperscript{1}, Naheem Noah\textsuperscript{1}, Sanchari Das\textsuperscript{1,2}} 
    \IEEEauthorblockA{
        \textsuperscript{1}\textit{Department of Computer Science, University of Denver, USA}\\
        \underline{\{muskan.fatima, naheem.noah, sanchari.das\}@du.edu}
    }
    % \IEEEauthorblockA{
    %     \textsuperscript{2}\textit{Information Sciences and Technology Department, George Mason University, USA}\\
    %     \underline{{sdas35}@gmu.edu}
    % }
}

\IEEEoverridecommandlockouts
\makeatletter\def\@IEEEpubidpullup{6.5\baselineskip}\makeatother
\IEEEpubid{\parbox{\columnwidth}{
    BuildSEC'24 Building a Secure \& Empowered Cyberspace\\
    19-21 December 2024, New Delhi, India\\
    % ISBN 1-891562-93-2\\
    % https://dx.doi.org/10.14722/ndss.2025.23xxx\\
    https://www.buildsec.org/\\
}
\hspace{\columnsep}\makebox[\columnwidth]{}}

% Make the title area
\maketitle

\begin{abstract}
The rise of online dating apps has transformed how individuals connect and seek companionship, with an increase in usage among older adults. While these platforms offer opportunities for emotional support and social connection, they also present significant challenges, including a concerning trend of online dating scams targeting this demographic. To address these issues, we conducted a semi-structured interview focused on the online dating experiences of older adults ($65+$). Initially, we conducted a pre-screening survey, followed by focused semi-structured interviews with $11$ of the selected older adults. Through this study, we investigate older adults' security and privacy concerns, the significance of design elements and accessibility, and identify areas needing improvement. Our findings reveal challenges such as deceptive practices, including catfishing and fraud, concerns over disclosing sensitive information, non-inclusive app design features, and the need for more informative visualization of match requests. We offer recommendations for enhanced identity verification, inclusive privacy controls by app developers, and increased digital literacy efforts to enable older adults to navigate these platforms safely and confidently.
\end{abstract}


\section{Introduction}
In today's digital age, online dating apps have become ubiquitous, bridging gaps across all age groups, cultures, and backgrounds~\cite{Bonilla-Zorita2021}. These virtual spaces have become crucial for providing emotional support and connection, particularly during the COVID-19 pandemic, which has exacerbated social isolation and loneliness~\cite {cairns2020covid,joshi2020substituting,gupta2024really}. Online dating apps streamline the matchmaking process and simplify the search for potential matches. For many users, these platforms bring companionship, emotional sustenance, and a renewed sense of belonging~\cite{BenefitOnlineDating,noah2024aging}. The widespread use of such technology in daily life has transformed how relationships begin and develop in the modern era~\cite{gupta2024critical}.

Online dating has also become popular among older adults (\textit{65+}). A 2023 Pew Research Center study revealed that 12$\%$ of seniors aged 70 and above have used a dating site or app, showing an increase from previous years~\cite{vogels2023key,noah2023safe}. These apps have also shown to be particularly appealing for divorced individuals or those widowed later in life seeking new relationships~\cite{viveros2018pros} thus often alluring older adults. Despite the advantages and widespread use of online dating, it remains susceptible to scams. Online dating scams present a significant threat to users, particularly older adults. In the context of this study, we define ``security risks'' as encompassing both technical security (such as data protection and privacy) and personal security (including protection from fraud, deception, and emotional harm). It is important to note that while many of these risks, particularly dating fraud, primarily involve a breach of trust, they also present significant security concerns for users, especially older adults. According to a $2023$ report by the Federal Trade Commission (FTC), romance scams led to reported losses amounting to $1.3$ billion, with adults aged 60 and above accounting for an alarming one-third of these victims~\cite{viveros2023top}. Scammers frequently target this demographic for various reasons, including perceived loneliness, social isolation, and a misconception that they are less adept with technology~\cite{ray2021older}. They typically create elaborate fake profiles and use stolen photographs to lure victims, often employing tactics such as excessive affection early on and urgent, emotionally charged requests for money or sensitive financial information. These deceptive strategies exploit the emotional and social vulnerabilities of older adults, leading to financial and emotional harm.

Although online dating scams are prevalent, this research area is severely understudied~\cite{noah2024evaluating}, particularly concerning older adults. We studied to understand older adults' perspectives on online dating to address this gap. We administered a pre-screening survey, and semi-structured interviews with $11$ participants. Our study examined the security and privacy challenges faced by older adults aged $65$ and above on online dating platforms and investigated the role of accessibility and design elements in these challenges. We aimed to answer the following research questions:

\begin{itemize}
\item \textbf{RQ1}: What are the most common privacy and security issues faced by older adults using online dating applications?
\item \textbf{RQ2}: How do the design and usability of online dating apps impact older adults' privacy and security? Are there particular design elements that help or hinder their safe use of these apps?
\item \textbf{RQ3}: What features or improvements do older adults need to better protect their security and privacy on dating apps?
\end{itemize}

\noindent \textbf{Contribution}: This research makes a significant contribution to the field by providing an analysis of how older adults aged $65$ and above interact with online dating platforms. This area is severely understudied, likely due to the taboo nature of online dating and the potential stigma associated with it. By focusing on both the emotional and technical challenges faced by this group, our findings emphasize the urgent need for improvements in app design to enhance user safety and privacy. This contribution is particularly valuable as it not only amplifies the voices of an often overlooked demographic but also lays a foundational basis for future research to build upon. Employing a user-centered approach, we uncover critical insights into the unique needs of older adults in the digital dating world, thereby informing better practices and policies for app developers and policymakers alike.

\section{Related Work}
The intersection of online dating, older adults, and digital technology presents a complex and evolving landscape that spans multiple research domains~\cite{das2019towards,das2020non,das2019don,zezulak2023sok,saka2023safeguarding}. 

\subsection{Online Dating Experiences of Older Adults}
A recent study conducted by SSRS reveals that over one-third (37\%) of American adults have experience with online dating platforms, having used either a dating website or mobile application at least once. Furthermore, the research indicates that 7\% of U.S. adults are actively engaged in online dating, currently using such sites or apps~\cite{ssrs_online_dating_2024}. Davis and Fingerman examined the profile content of older adults in online dating, finding that older adults often emphasize positive self-representation and established social connections~\cite{geronb}. Similarly, Griffin and Fingerman studied online dating profiles of older adults seeking same and cross-sex relationships, revealing nuanced variations in self-presentation strategies. They found that older adults seeking same-sex relationships often emphasized romantic elements more than their heterosexual peers~\cite{Griffincrossex}. Wada et al. analyzed older adults' self-representations in online dating profiles, identifying five primary themes: an active lifestyle, physical and intellectual engagement, ongoing productivity, and descriptions of themselves as happy, fun-loving, and humorous~\cite{wada2019busy}. This aligns with Watson and Stelle's work on self-presentation in online dating advertisements of older adults, which highlighted the importance of positive self-portrayal~\cite{watson2021love}.

Further research by Gewirtz-Meydan and Ayalon explored the motivations of older adults for using online dating sites, finding that they sought companionship, intimacy, and a sense of belonging~\cite{gewirtz2018forever}. Additionally, Marston et al. examined the impact of technology on the romantic lives of older adults, noting that online dating platforms can provide opportunities for social connection and emotional fulfillment~\cite{marston2020ok}.

\subsection{Privacy and Security Concerns for Older Adults}
Privacy and security are significant concerns for older adults using online dating platforms. A $2023$ report by the Federal Trade Commission (FTC) revealed that romance scams led to reported losses amounting to 1.3 billion, with adults aged 60 and above accounting for one-third of these victims~\cite{viveros2023top}. Scammers frequently target this demographic due to perceived loneliness, social isolation, and a misconception that they are less adept with technology~\cite{ray2021older}. Ray et al. explored why older adults don't use password managers, providing insights into their general online security practices. Their findings suggest that older adults may be more vulnerable to security threats due to unfamiliarity with advanced security measures~\cite{ray2021older}. Vitak and Shilton discussed trust, privacy, security, and accessibility considerations when conducting mobile technology research with older adults. Their recommendations include meticulous informed consent procedures and stringent data protection measures~\cite{vitak2020trust}. Similarly, Wang and Price examined accessible privacy, emphasizing the need for privacy controls that are easily navigable by older adults~\cite{wang2022accessible}. This aligns with Shetty et al.'s study on the security of Android dating apps, which highlighted the importance of robust security features in dating applications~\cite{shetty2017you}.

Recent work by Frik et al. explored privacy and security decision-making in older adults, finding that they often rely on different cues and heuristics compared to younger users~\cite{frik2019privacy}. Additionally, Nicholson et al. examined the impact of age on susceptibility to phishing attacks, providing insights into how older adults might be targeted by online dating scams~\cite{nicholson2019if}. While previous studies have examined general online security behaviors of older adults, our work provides targeted insights into the specific risks and challenges faced by this demographic when seeking romantic connections online. 

\subsection{Usability and Accessibility in Mobile Apps for Older Users}
Díaz-Bossini and Moreno discussed the barriers posed by declining vision, reduced dexterity, and cognitive limitations in mobile interfaces for older people~\cite{diaz2014accessibility}. To address these challenges, several studies advocate for universally accessible designs~\cite{elguera2019elderly,kascak2014integrating,wong2018usability}. Kascak et al. and Harrington et al. emphasized the need for clear layouts, large fonts, and simplified navigation to enhance the user experience for older adults~\cite{kascak2014integrating,harrington2017universally}.].
De Lima Salgado et al. provided a practical guide on UX, usability, and accessibility evaluation for older adults using mobile applications~\cite{de2019startup}. Their work complements Wildenbos et al.'s research on barriers influencing mobile health usability for older adults, which identified age-related barriers affecting mobile health usability~\cite{wildenbos2018aging}. Bao and Pan studied the acceptability of mobile payments by older adults in China, highlighting the importance of considering cultural and technological familiarity in designing for older users~\cite{bao2021study}.

Recent work by Nurgalieva et al. also explored the design of mobile banking apps for older adults, emphasizing the importance of trust-building features and clear information presentation~\cite{nurgalieva2019information}. Similarly, Petrovčič et al. investigated the adoption of mobile health apps among older adults, finding that perceived usefulness and ease of use were key factors in adoption~\cite{petrovvcivc2018smart}. While previous research has focused on general mobile app usability for older users, our study provides insights into the particular needs and preferences of this demographic when navigating the complex social and emotional landscape of online dating platforms. Our analysis contributes uniquely to this field by examining the specific usability and accessibility challenges faced by older adults in the context of online dating apps.

\section{Method}
% To investigate the security and privacy issues faced by older adults using online dating applications, we employed a user-centered approach consisting of a pre-screening survey and semi-structured interviews.

\subsection{Recruitment and Pre-Screening Survey}
We initiated our study by distributing flyers through approved bulletin boards, social media networks, and local community spaces. Additionally, we collaborated with several healthy aging institutions, including the Knoebel Institute for Healthy Aging (KIHA) at the University of Denver, to advertise our study in their newsletters and to use appropriate recruitment terminologies for the target population. Interested participants with prior experience using dating applications filled out a screening recruitment form on Qualtrics. We received $180$ responses. The survey included consent, pre-screening information, online dating experiences, cybersecurity knowledge, and demographics. We explored areas such as participants' motivations for using dating apps, safety and privacy concerns, experiences with scams or deception, and general computer security practices. To ensure the quality of our sample, all authors conducted a detailed evaluation of the survey responses and ultimately selected $11$ participants who consented to share their experiences with online dating applications. The selection criteria required participants to be aged $65$+ and reside within the United States.
 
\subsection{Interview Process}
We conducted interviews with participants either in person or via Zoom, with all preferring the latter method. The first author led all Zoom interviews and provided each participant with a \$20 electronic gift card for hour-long sessions. After each interview, we gathered feedback to clarify the interview protocol, especially technical terms, refining our semi-structured guide through this iterative process. After four interviews, we modified our protocol to explore participants' dating preferences, reasons for deleting apps, and identity verification methods, enhancing data collection by addressing previously overlooked areas. Our final sample comprised eight men and three women, with ten identifying as heterosexual and one as part of the LGBTQIA+ community. All participants, except one (P10), held a four-year degree and were employed full-time. We conducted this study between March 2023 and April 2024.

\subsection{Ethical Considerations}
The ethics committee review board of the primary research organization approved the study. Informed consent was obtained from all participants, and they were informed that they could opt out at any point. Participants were also informed that they might share personal, sensitive, and identifiable information during the study. To ensure anonymity, we took measures to anonymize any such information. We avoided email communication regarding interview details to uphold privacy, and we deleted any personally identifiable information promptly after analyzing the interviews. All data were stored in encrypted, password-protected cloud storage accessible only to the authors. 

\subsection{Data Analysis}
We employed thematic analysis to understand the security, privacy, usability, and accessibility challenges faced by older adults on online dating apps~\cite{braun2021can}. Our interviews aimed to capture the experiences, needs, and preferences of older adults using dating apps. To ensure data homogeneity, we restricted our participant pool to individuals residing in the United States at the time of survey completion. Participants were asked to share their positive and negative experiences with dating apps, indicate their preferred platforms, and evaluate the design features they found either unique or challenging. Additionally, we explored their perceptions of the ease of managing privacy settings within these apps.

Using an inductive coding approach and structural coding, we identified themes based on our research questions. The first author collected all the codes and collaborated with the second author using Collaborative Qualitative Analysis (CQA)~\cite{richards2018practical} to finalize a codebook (Table~\ref{tab:codes}). We ensured the validity of our findings by meeting multiple times to discuss codes, themes, and analyses with the entire research team. After analyzing the interviews, we identified themes related to older adults' online dating experiences and developed design and accessibility recommendations for dating applications. We discuss these findings in subsequent sections.

\begin{table}[ht]
\centering
\begin{tabular}{|l|p{6.3cm}|p{5cm}|}
\hline
\textbf{Code} & \textbf{Description} \\ \hline
% P01-P011 & Participant Number \\ \hline
Ar01-Ar18 & App Recommendations for Enhanced Privacy and Security \\ \hline
% Af01-Af10 & App Features Recommended by Participants \\ \hline
Ps01-Ps08 & Personal and Social Impact on Participants \\ \hline
Mp01-Mp03 & Managing Privacy Settings \\ \hline
Pc01-Pc05 & Privacy and Security Concerns \\ \hline
Pi01-Pi17 & Personal Information \\ \hline
Cn01-Cn16 & Challenges and Negative Aspects \\ \hline
Mv01-Mv10 & Motivation \\ \hline
M01-M04 & Meeting Connections \\ \hline
% A01-A12 & Dating Apps Used by Participants \\ \hline
Pm01-Pm15 & Privacy and Security Measures Taken by Participants \\ \hline
Av01-Av06 & Authenticity Verification \\ \hline
\end{tabular}
\caption{Description of Codes Used in the Study}
\label{tab:codes}
\end{table}

\section{Findings and Discussions}
Our analysis of interviews with $11$ older adults (65+) using online dating platforms revealed key insights into their experiences, challenges, and concerns. This section presents our findings and discussions on designing online dating apps.

\subsection{Dating App Usage and Connections}
\subsubsection{Dating App Usage}
Our exploration of the dating app landscape among older adults revealed a diverse range of preferences and experiences. The $11$ participants in our study collectively used a total of $12$ different dating applications, which we have tagged as \textit{A1} through \textit{A12}. The number of apps used per participant varied, with  \textit{P05}, having explored four different platforms. On the other hand, six participants (\textit{P01}, \textit{P03}, \textit{P04}, \textit{P06}, \textit{P07}, and \textit{P10}) chose to stick with a single app. Among the apps mentioned, Tinder (\textit{A2}) emerged as the most popular choice, with four participants having used it at some point. This finding aligns with a recent Pew Research study, which found that $19\%$ of online daters aged $50$ to $59$ have used Tinder~\cite{pewresearch2023}. Bumble (\textit{A3}) and PlentyOfFish (\textit{A5}) tied for second place, each being used by three participants. Several other apps made single appearances in our study, including Bloom Meet Singles (\textit{A1}), Matchmakers (\textit{A4}), Hinge (\textit{A6}), OkCupid (\textit{A7}), Zoosk (\textit{A8}), InternationalCupid (\textit{A9}), Christian Mingle (\textit{A10}), and Badoo (\textit{A12}). This diversity highlights the fact that older adults have varying preferences and may be drawn to niche apps that cater to specific interests, religions, or geographic locations. Interestingly, two participants reported using Facebook Dating (\textit{A11}). Although known as a social networking site, Facebook's dating feature has seen increasing adoption among users seeking romantic connections~\cite{facebook_dating}.

% \subsubsection{Participant Demographics}
% While we were able to interview 12 participants, only 5 provided voluntary demographic data which we are reporting here:
% \begin{table}[h!]
% \centering
% \begin{tabular}{| m{3cm} | m{3cm} |}
% \hline
% \textbf{Demographic} & \textbf{Number of Participants} \\ \hline
% \textbf{Race} & \\ \hline
% African American & 5 \\ \hline
% \textbf{Gender} & \\ \hline
% Male & 4 \\ \hline
% Female & 1 \\ \hline
% \textbf{Highest Level of Education Completed} & \\ \hline
% 2 year degree & 1 \\ \hline
% Professional degree & 1 \\ \hline
% 4 year college & 3 \\ \hline
% \textbf{Current Employment Status} & \\ \hline
% Employed part-time & 3 \\ \hline
% Employed full time & 2 \\ \hline
% \textbf{Current Annual Household Income} & \\ \hline
% \$10k-19k & 1 \\ \hline
% \$20k-30k & 1 \\ \hline
% \$40k-50k & 1 \\ \hline
% \$50k-60k & 1 \\ \hline
% \$100k-150k & 1 \\ \hline
% \textbf{Current Relationship Status} & \\ \hline
% Married & 3 \\ \hline
% Separated & 2 \\ \hline
% \end{tabular}
% \caption{Demographic Information of Participants}
% \label{tab:demographics}
% \end{table}

\subsubsection{Dating App Deletion - RQ1}
%RQ1
App deletion was a common occurrence among the participants in our study, with over six individuals deleting at least one dating app at some point during their online dating journey. The reasons for deletion varied, ranging from disappointment with matches to experiences of catfishing and fraud. \textit{P05} deleted the highest number of apps, removing both Tinder (\textit{A2}) and Bumble (\textit{A3}) from their device. 
% The table~\ref{tab:app-usage} attached in the appendix provides information on apps used and deleted by each participant. 
% Other participants, including \textit{P04}, \textit{P06}, \textit{P09}, \textit{P10}, and \textit{P11}, also deleted apps. Interestingly, out of the seven apps that were deleted, Tinder (\textit{A2}), Bumble (\textit{A3}), and PlentyOfFish (\textit{A5}) were each deleted twice, while InternationalCupid (\textit{A9}) was deleted by the single participant who used it. 
Some participants, 
like \textit{P09}, deleted apps like Bumble (\textit{A3}) after experiencing catfishing, explaining, \\
\begin{quote}
    \textit{``Yeah, and how I did it once [was to] delete a dating app. [And for that reason,] it was because of the catfish the first time. That's when I deleted the app.''}
\end{quote}

\subsection{Motivations \& Goals}
When it comes to choosing a dating app, older adults tend to be more intentional and purposeful compared to younger generations. Our study identified ten distinct motivations that drive older adults to join dating apps, each reflecting their unique needs, desires, and life circumstances. The most common reason, cited by four participants, was the desire to meet new people and socialize. 
% (\textit{Mv01}). 
For example, \textit{P02} expressed the difficulty of meeting potential partners through traditional means, stating, 
\begin{quote}
    \textit{``I think what really inspired me to start using online dating apps, particularly, is that sometimes it's very difficult to, you know, [to] meet someone, first of all...''}
\end{quote}

This sentiment highlights the appeal of dating apps as a convenient and effective way to connect with others. Interestingly, three participants joined dating apps based on referrals from friends or family,
% (\textit{Mv08})
 suggesting that trust and personality play a significant role in the decision-making process for some older adults. Other motivations included the need to find companionship or emotional support following a recent breakup,
% (\textit{Mv02})
confidence,
% (\textit{Mv03})
 frustration with people,
% (\textit{Mv04})
 the quest to find a soulmate,
% (\textit{Mv06})
the desire for a casual or short-term relationship,
% (\textit{Mv07})
 lack of options,
% (\textit{Mv09})
 and the ability to expand location options to find matches.
% (\textit{Mv10})

\subsection{Personal and Social Impact - RQ1}
%RQ1
Online dating platforms have profoundly influenced the social lives and personal relationships of older adults, as revealed by our participants' experiences. Most participants reported that using these apps has helped improve their communication skills, 
% (\textit{Ps02}), 
with four individuals specifically highlighting this benefit. \textit{P03} described the positive influence of online dating on their communication abilities and social interactions:
\begin{quote}
    \textit{``I think there's positive role and negative impact about it. And it has been able to do [help with] my communication skills. You know, I have this confidence of you know, starting up a conversation and wanting to meet people [and] wanting to know people more, and of course, making friends, and see what works out for me, and what could make me feel better''}.
\end{quote}
However, the benefits of online dating extend beyond communication skills. \textit{P06} shared a comprehensive account of how using these platforms has positively affected various aspects of their life, including improved communication, 
% (\textit{Ps02}), 
increased confidence, 
% (\textit{Ps03}), 
diversified relationships, 
% (\textit{Ps04}), 
and enhanced professional networking abilities. 
% (\textit{Ps05}). 
However, one participant \textit{P02} expressed concern that relying on online dating applications might have negatively impacted their face-to-face communication skills. 

\subsection{User Experience \& Usability - RQ2 \& RQ3}
%RQ2 AND RQ3
\subsubsection{App Design Features}
When asked about the most useful features of the dating apps they have used, participants highlighted several key components that enhanced their overall experience. User-friendliness 
% (\textit{Af07}) 
emerged as the most critical aspect, with four participants emphasizing its importance. \textit{P05} noted that the apps are designed to be accessible to users with minimal education, stating,
\begin{quote}
\textit{``I think all of the apps are not complicated. They're made to be user-friendly. [...] In fact, you just need your kindergarten and education to be able to use most of them. That means everybody. Once you can read and write, know how to pronounce things. you'll be able to use them. I didn't find any difficulties using them. but I also would say being too simple is the reason why most of the apps are abused''}\\
\end{quote}
However, \textit{P05} also cautioned that this simplicity could be dangerous, as it may contribute to the apps being abused. Ease of navigation 
% (\textit{Af06}) 
was another critical factor mentioned by three participants. A participant (\textit{P08}) noted using Messenger, a Facebook-based messaging app, to meet new people in addition to using other dating apps. \textit{P08} specifically praised the user-friendly interface of Facebook's Messenger feature (\textit{A11}), stating:

\begin{quote}
\textit{``Messenger is very easy to use. [...] I think Messenger is the most helpful for me.''}
\end{quote}

Participants also mentioned other design features that they found interesting and helpful, such as the search icon, the availability of swiping, and the presence of voice/video calls. 

\subsubsection{App Recommendations for Enhanced Privacy and Security}
%RQ3
Participants collectively identified over $18$ unique features that they believe could significantly improve the functionality and user-friendliness of dating applications. Among these recommendations, improved match settings 
% \textit{(Ar02)} 
emerged as the most frequently cited feature, with six participants emphasizing its importance. For instance, \textit{P08} expressed a desire for greater control over who can initiate contact, stating:
\textit{
\begin{quote}
    ``[..] People should not be given the opportunity to send [friend/match] requests to everyone online. [people] should be given some instruction[instructions]. [paraphrased and verified - I think there should be instructions before you send a request to someone, and  become their friend'].''
\end{quote}
}
Another prominent recommendation, voiced by five participants, was the implementation of improved identity verification measures. 
% \textit{(Ar11)}. 
Furthermore, two participants expressed a desire for greater app policy transparency,  
% \textit{(Ar04)}, 
advocating for clear and accessible information regarding the platform's data handling practices and user protections. Complementing this, two other participants advocated for enhanced user control over personal information, suggesting implementing features that allow for locking or restricting access to certain data. 
% \textit{(Ar10)}. 
One of the participants also suggested educating all users about the use of dating apps and how to use them ethically. Moreover, another user also suggested collaboration of dating applications with law enforcement so individuals who abuse dating apps can easily be reported. 

\subsection{Privacy, Security \& Information Sharing - RQ1 \& RQ3}
%RQ1
\subsubsection{Sharing Personal Information}
Understanding the information older adults are willing to disclose on online dating platforms and the details they prefer to keep private is crucial in addressing privacy and security concerns. Most participants opted to share their names 
% \textit{(Pi02)} 
and hobbies/preferences 
% \textit{(Pi04)}, 
with four participants indicating their willingness to disclose each of these categories, respectively. While two participants expressed openness to sharing their ages, 
% \textit{(Pi05)}, 
one participant each chose to reveal their alternative contact information,
% \textit{(Pi01)}, 
state of residence, 
% \textit{(Pi03)}, 
race, 
% \textit{(Pi06)}, 
educational status, 
% \textit{(Pi07)}, 
photos, 
% \textit{(Pi08)}, 
and preferred meet-up locations. 
% \textit{(Pi09)}. 
\textit{P01} shared their approach to information disclosure:\\
\begin{quote}
\textit{``I will always share my, you know, my alternative contact, not my personal contact. Yeah, but my name? Yes, I share my name. My, you know, my state where we [live], but I don't, I know, put in my address, my residential address, to, you know, to be very on the very steps. That's what I do.''}
\end{quote}

Similarly, \textit{P04} expressed willingness to share information such as hobbies but exercised caution when it came to disclosing personal details like addresses and ages:

\begin{quote}
\textit{``Okay, my [profile includes] hobbies, [things like that], but not my personal information like my address because maybe someone [could be] looking for people's addresses. And contacts, too. [They might] interact with you [or] interfere with [your activities]. Yeah. So the only things I could [be concerned about sharing are] my age, my pictures, [and] the hobbies [and] things I like [or] dislike.''}
\end{quote}

Conversely, most participants preferred not to share personal contact information, 
% \textit{(Pi10)}, 
residential addresses, 
% \textit{(Pi11)}, 
work addresses, 
% \textit{(Pi12)}, 
family details, 
% \textit{(Pi13)}, 
banking information, 
% \textit{(Pi14)}, 
their race, 
% \textit{(Pi15)}, 
institutional affiliations, 
% \textit{(Pi16)}, 
and their exact names. 
% \textit{(Pi17)}. 
The categories of personal contact information 
% \textit{(Pi10)} 
and residential addresses 
% \textit{(Pi11)} 
emerged as the most guarded, with five participants each indicating their reluctance to disclose these details. Family details 
% \textit{(Pi13)} 
were cited by four participants, while work addresses 
% \textit{(Pi12)} 
and banking information 
% \textit{(Pi14)} 
were mentioned by three participants each as information they would not share. Each participant mentioned the remaining categories.

\subsubsection{User Verification}
Verifying the authenticity of potential matches is a crucial aspect of ensuring a safe and trustworthy online dating experience. Our study sought to understand the various methods employed by older adults to confirm the legitimacy of their connections on these platforms. Surprisingly, most of our participants relied heavily on profile or picture exploration 
% \textit{(Av05)} 
as their primary verification strategy, with four participants mentioning this approach. Another commonly cited verification method was requesting a phone call or directly calling the person of interest, 
% \textit{(Av01)}, 
with three participants endorsing this practice. In addition to these dominant verification strategies, participants also reported employing other methods to ensure the legitimacy of their matches. These included arranging in-person meetings,
% \textit{(Av02)}, 
relying on verified account features provided by the dating platforms themselves, 
% \textit{(Av04)}, 
and assessing the level of communication and consistency exhibited by their potential partners 
% \textit{(Av06)}.

\subsection{Privacy Challenges \& Negative Aspects - RQ1 \& RQ3}
%RQ1
The participants reported several negative aspects. Being victims of financial fraud is the most prevalent (\textit{P06, P07, P08, P09}). Participants mentioned being asked for money in return for companionship.  Unsurprisingly, the phenomenon of catfishing, or the use of fake profiles by individuals, 
% \textit{(Cn12)}, 
emerged as a prevalent concern, with four participants indicating they had fallen victim to such deceptive practices. \textit{P09} vividly recounted their experience of being catfished multiple times, on Facebook \textit{(A11)}:\\
\begin{quote}
\textit{``Yeah, I think it's been [twice]. The third time being [when I was] catfished by people. [It] was actually on Facebook. You know the way you can [connect] with a girl from the comments and you decide to DM [her] trying to get to know [her], and she just [gives off a] vibe that it's something you've been looking for. Like for the first 2 to 3 days, she pretends to be what you want. You know, and [then] she starts asking for money for [various] things. And because you're trying to impress [her], you just [give her] money. And also, [I have] been [fooled] by a lady who [asked me to send] money for an Uber, [instead of me paying directly], and she never showed up.''}
\end{quote}

In addition to catfishing and fraud, three participants highlighted the issue of unmatching identity, 
% \textit{(Cn03)}, 
where users present inaccurate or outdated photos, leading to disappointment or misrepresentation. Furthermore, three participants cited dating apps as a ``time waste''. 
% \textit{(Cn13)}. 
Other challenges mentioned by participants included reduced self-esteem, 
\textit{(Cn01)}, 
addiction, 
% \textit{(Cn02)}, 
sexual abuse, 
% \textit{(Cn03)}, 
physical engagement, 
% \textit{(Cn04)}, 
stalking, 
% \textit{(Cn05)}, 
naked displays, 
% \textit{(Cn06)}, 
lies, 
% \textit{(Cn07)}, 
sex work, 
% \textit{(Cn08)}, 
lack of information, 
% \textit{(Cn10)}, 
disappointed outcomes, 
% \textit{(Cn11)}, 
safety concerns, 
% \textit{(Cn14)}, 
and stress.
% \textit{(Cn16)}.

\subsubsection{Privacy and Security Measures for Meeting Matches}
Most participants implemented various security and privacy measures before meeting their matches in person. The predominant precaution involved arranging to meet in a public location, ensuring a safe and neutral environment \textit{P04, P06, P10}. Additionally, some participants indicated a preference for organizing casual outings or informal gatherings with their matches to establish a comfortable and low-pressure atmosphere. These measures reflect a broader awareness and proactive approach to personal safety and privacy in the context of online dating.

\section{Implications}
This study showcases the intricate relationship between the motivations, experiences, and concerns of older adults on online dating platforms. While participants actively sought connection and companionship, they faced significant challenges and vulnerabilities, particularly in the form of deceptive practices such as catfishing and dating fraud. These negative experiences not only eroded trust in these platforms but also led some participants to delete their dating apps altogether. This underscores the urgent need for enhanced security measures, including more rigorous match settings and improved identity verification processes, to protect users from malicious actors. Additionally, our findings highlight the critical need for balancing information sharing with privacy protection. Although participants were willing to share basic details such as names and hobbies, they remained cautious about divulging sensitive personal information, reflecting a heightened awareness of security risks.
% This cautious approach indicates a clear demand for privacy-conscious design in online dating platforms, where users can feel secure in their interactions without compromising their personal information.
%This behavior highlights potential shortcomings in current privacy practices and the necessity for enhanced user education. Moreover, the study reveals a reliance on visual cues and profile checks to verify match legitimacy—an approach that fails to counteract the advanced deceit tactics of malicious actors.

\subsection{Advanced Identity Verification}
Our study reveals significant challenges with scams and deceptive practices, such as catfishing, which disproportionately affect older adults. Several participants (\textit{P06, P07, P08, P09}) reported falling victim to financial fraud, often being solicited for money in exchange for companionship. Additionally, four participants encountered catfishing, highlighting the need for robust measures to combat these risks. To address these concerns, we recommend implementing advanced identity verification systems. Real-time video verification during account setup can substantially reduce the prevalence of fake profiles\cite{centelles2021examination,zhao2014quasi}, particularly addressing older adults' tendency to rely on profile pictures for initial trust-building. 

Integrating two-factor authentication at login adds a crucial security layer~\cite{das2020don}, especially for older adults who may be less familiar with advanced security practices. Where permissible and compliant with privacy laws, linking profile verification with National ID systems offers a reliable method for confirming user identities~\cite{mcgrath2016identity}. Additionally, a peer vouching system, where trusted long-term users vouch for new users they know in real life, could foster a network of trust—especially valuable for older adults who often join online dating through personal recommendations. Implementing a tiered, progressive verification system, where users can verify additional aspects of their identity (e.g., profession, education) to earn ``trusted" badges, addresses concerns about the authenticity of profile information. Finally, prominently displaying verification status with clear and easily understandable icons or labels caters to older adults' preference for straightforward interface elements, enhancing their confidence and safety on these platforms.

\subsection{Enhanced Privacy Controls}
Our findings reveal that older adults using dating applications value their privacy and seek greater control over the information they share.
% (\textit{Pi10, Pi11, Pi13}). 
To meet this need, dating platforms should offer enhanced and user-friendly privacy controls that allow individuals to manage who can view their profiles, thereby limiting exposure to only those they choose~\cite{shetty2017you}. This can be challenging given the nature of the functionality of these platforms for connecting strangers. While some participants (\textit{P02, P09, P10, P11}) found managing privacy settings straightforward, others expressed a preference for clearer and simpler options. Therefore, designing privacy settings with clarity and simplicity is essential, and should be supported by step-by-step guides or tutorial videos~\cite{wang2022accessible}.

Implementing granular controls that allow users to selectively share information based on their level of interaction with others can further address concerns about sharing sensitive details, such as addresses 
% (\textit{Pi11}) 
or family information. 
% (\textit{Pi13}). 
% Regular privacy audits can help users stay informed and in control of their information-sharing practices~\cite{shelton2010case}. 
Finally, features like ``incognito mode'' can enable users to browse matches discreetly, offering protection against unsolicited contacts.
% (\textit{Cn05}). 
% Finally, providing transparent information about data collection and usage fosters trust and empowers users to make informed decisions, addressing participants' desire for greater clarity in-app policies. 
% (\textit{Ar04}).
% Given the particular vulnerability of older adults to privacy invasions, dating applications must provide enhanced and easily navigable privacy controls. Users should be able to adjust who can view their profiles, effectively limiting exposure to only those individuals they wish to engage with \cite{8207632}. It is crucial that privacy settings are straightforward and accessible, supplemented by supports such as step-by-step guides or tutorial videos to aid those less familiar with digital environments \cite{wang2022accessible}. Furthermore, regular privacy audits conducted by the application can help maintain user awareness and control over their personal information sharing \cite{shelton2010case}.

\subsection{Educational Initiatives}
Enhancing digital literacy plays a key role in empowering older adults to navigate online dating platforms confidently and safely. Our findings highlight varying levels of technological proficiency among participants, underscoring the importance of accessible education. Collaborative online safety workshops, organized with community centers and senior organizations, can offer valuable guidance on recognizing and avoiding online scams~\cite{ball2002effects}. Developing clear and accessible resources, such as instructional videos and printable guides, will support older adults in mastering online interactions~\cite{reuter2021content}, especially for those who benefit from additional guidance with privacy settings. Implementing a peer mentoring system, where tech-savvy seniors can guide their peers, can enhance both competence and confidence in using app features. 
% In-app tutorials and periodic safety reminders, tailored specifically for older users, can further reinforce safe practices. These initiatives should also address the emotional aspects of online dating, equipping older adults with the tools to identify and avoid manipulative tactics in romance scams. 
Encouraging family involvement, as suggested by the influence of referrals 
% (\textit{Mv08}) 
in our findings, can provide additional support and foster a more secure online dating experience for older adults.

% Enhancing digital literacy is crucial for helping older adults navigate online dating safely. Collaborative online safety workshops with community centers and senior organizations can provide essential education on recognizing and avoiding online scams \cite{kulinski2017advancing,ball2002effects}. Furthermore, developing clear and accessible resources, such as instructional videos and printable guides, will aid older adults in understanding how to interact safely online \cite{reuter2021content}. Implementing a peer mentoring system, where tech-savvy seniors guide their peers through dating platforms, can also significantly boost their competence and confidence.

\subsection{Policy Recommendations}
Policymakers should implement regulations that require dating platforms to adopt rigorous identity verification processes and maintain transparent privacy policies, addressing concerns about deceptive practices and information security raised by our participants. Legislation mandating regular security audits and strict data protection compliance will ensure that dating platforms meet high safety standards, especially in protecting users from online scams (\textit{P06, P07, P08, P09} experienced fraud). Policies should also mandate that dating apps provide clear and accessible information about data handling practices, responding to the widespread call for greater transparency in app policies.
% (\textit{Ar04}). 
Additionally, enforcing age-appropriate design principles in dating apps is crucial. This includes considering older users' needs in interface design, accessibility, and privacy controls, aligning with feedback on usability challenges. Furthermore, policymakers should expand support for community-based digital literacy programs that equip older adults with the necessary skills to navigate online environments securely. Establishing a dedicated helpline for older adults facing online dating issues could provide timely assistance and guidance. 
\section{Limitations and Future Work}
Our study offers insights into the security and privacy issues older adults face with online dating apps. However, the small sample size may limit the generalizability of our findings. Additionally, our self-reported data may introduce recall and social desirability biases, which are common in qualitative research. To address these limitations, we plan to expand our sample size and include participants from diverse cultural and socio-economic backgrounds in future studies. We aim to explore the experiences of older adults who identify as part of the LGBTQIA+ community, as they may encounter unique challenges not fully addressed in this study. We also intend to use longitudinal designs to assess the effectiveness of security measures and privacy controls over time. 

\section{Conclusion}
This study has uncovered significant potential for improving social connections among older adults through dating apps, while also addressing crucial privacy and security concerns. To do this, we pre-screened participants and conducted semi-structured interviews with $11$ older adults, gaining valuable insights into their interactions with online dating platforms. Participants approached dating apps with intentionality and mindfulness, reflecting a wide range of preferences and experiences. However, more than half of them eventually deleted an app, driven by concerns about privacy, security, or usability. Although they achieved some progress in communication skills and socializing, participants remained worried about information sharing and the non-inclusive design of these platforms. Negative experiences such as catfishing and fraud emphasized the urgent need for enhanced security and privacy measures. To improve the safety and inclusivity of these platforms for older adults, we recommend incorporating identity verification features to prevent fraud, enhance app accessibility, and provide clearer information on privacy settings. Addressing these concerns will help create a more secure and welcoming environment for older adults on dating apps.
\section{Acknowledgement}
This work has been supported by the Partners in Scholarship (PinS) grant for the projects~\lq\lq Evaluating the Privacy and Security Perspective of Older Adults Towards Online Dating\rq\rq~ and~\lq\lq SafeHeart: Addressing Online Dating Privacy and Security Needs of the Older Adults\rq\rq~. We would also like to thank our participants for this study and acknowledge the Inclusive Security and Privacy-focused Innovative Research in Information Technology (InSPIRIT) Lab at the University of Denver for supporting this work. Any opinions, findings, conclusions, or recommendations expressed in this material are solely of the authors.
% \section{Conclusion}
% Studies on older adults’ experiences with dating apps is limited but has significant potential to enhance social connections while ensuring privacy and security. In our study, we conducted a pre-screening survey followed by semi-structured interviews with $11$ participants, which offered valuable insights into their use of online dating platforms. Our findings show that participants use dating apps with intentionality and mindfulness, revealing a diverse range of preferences and experiences. However, over half of the participants had deleted an app at some point due to concerns about privacy, security, or usability. Despite their goals of meeting new people and socializing, participants noted improvements in their communication skills but continued to express significant worries about information sharing and the non-inclusive design of these platforms. They also reported negative experiences, such as catfishing and fraud, underscoring the need for better security and privacy measures. Based on these findings, we recommend enhancing the inclusivity and accessibility of dating apps, implementing security features like identity verification to prevent fraud and catfishing, and providing clearer information on privacy settings to improve user security. Addressing these issues will help make online dating platforms safer and more welcoming for older adults.

% \balance
\bibliographystyle{plain}
\bibliography{references}

% \appendix
% \begin{table}[ht]
% \centering
% \caption{Participants' Dating App Usage}
% \label{tab:app-usage}
% \begin{tabular}{|c|p{3cm}|p{3cm}|}
% \hline
% \textbf{Participant} & \textbf{Apps Used} & \textbf{Apps Deleted} \\
% \hline
% P01 & Bloom Meet Singles (A1) & \\
% \hline
% P02 & Tinder (A2), Bumble (A3) & \\
% \hline
% P03 & Matchmakers (A4) & \\
% \hline
% P04 & Plenty Of Fish (A5) & Tinder (A2) \\
% \hline
% P05 & Bumble (A3), Hinge (A6), OkCupid (A7), Zoosk (A8) & Tinder (A2), Bumble (A3) \\
% \hline
% P06 & InternationalCupid (A9) & InternationalCupid (A9) \\
% \hline
% P07 & Christian Mingle (A10) & \\
% \hline
% P08 & Tinder (A2), Facebook (A11), Badoo (A12) & \\
% \hline
% P09 & Tinder (A2), Facebook (A11) & Bumble (A3) \\
% \hline
% P10 & Plenty Of Fish (A5) & Plenty Of Fish (A5) \\
% \hline
% P11 & Tinder (A2), Bumble (A3), Plenty Of Fish (A5) & Plenty Of Fish (A5) \\
% \hline
% \end{tabular}
% \end{table}
\end{document}